\documentclass[3p,twocolumn]{elsarticle}
\journal{Physics Letters B}

\usepackage{amsmath}
\usepackage{amssymb}
\usepackage{graphicx}
\usepackage{ulem}

\newcommand{\be}{\begin{equation}}
\newcommand{\ee}{\end{equation}}
\newcommand{\ba}{\begin{eqnarray}} 
\newcommand{\ea}{\end{eqnarray}} 
\newcommand{\nn}{\nonumber}
\DeclareMathAlphabet\mathbfcal{OMS}{cmsy}{b}{n} 

\usepackage[usenames,dvipsnames]{color}

\begin{document} 

\title{Jet quenching in glasma}

\author[bu,wi]{Margaret E. Carrington}
\author[ncbj]{Alina Czajka}
\author[ncbj,ujk]{Stanis\l aw Mr\' owczy\' nski}

\address[bu]{Department of Physics, Brandon University,
Brandon, Manitoba R7A 6A9, Canada}
\address[wi]{Winnipeg Institute for Theoretical Physics, Winnipeg, Manitoba, Canada}

\address[ncbj]{National Centre for Nuclear Research, ul. Pasteura 7,  PL-02-093  Warsaw, Poland}
\address[ujk]{Institute of Physics, Jan Kochanowski University, ul. Uniwersytecka 7, PL-25-406 Kielce, Poland}

\date{September 23, 2022}

\begin{abstract}

We discuss the transverse momentum broadening of hard probes traversing an evolving glasma, which is the earliest phase of the matter produced in relativistic heavy-ion collisions. The coefficient $\hat q$ is calculated using the Fokker-Planck equation, and an expansion in the proper time $\tau$ which is applied to describe the temporal evolution of the glasma. The correlators of the chromodynamic fields that determine the Fokker-Planck collision terms, which in turn provide $\hat q$, are computed to fifth order in $\tau$. The momentum broadening is shown to rapidly grow in time and reach a magnitude of several ${\rm GeV^2/fm}$. We show that the transient pre-equilibrium phase provides a contribution to the energy loss of hard probes which is comparable to that of the long lasting, hydrodynamically evolving, equilibrium phase. 

\end{abstract}

\maketitle

Hard probes -- high $p_\perp$ partons and heavy quarks -- are produced through interactions with large momentum transfer at the earliest phase of a heavy-ion collision at the Relativistic Heavy Ion Collider (RHIC) and the Large Hadron Collider (LHC). The hard probes propagate across the strongly interacting matter produced in the collision, and lose a substantial fraction of their initial energy. This is known as jet quenching. The phenomenon is treated as a signal of the formation of quark-gluon plasma because only the deconfined state of matter could be so opaque to color charges, see the review~\cite{Wiedemann:2009sh}. 

The matter produced in relativistic heavy-ion collisions evolves rapidly towards a locally equilibrated quark-gluon plasma which expands hydrodynamically and ultimately hadronizes. One expects that the final momentum spectra of hard probes are shaped in the long-lasting equilibrium phase which is relatively well understood. Effects of the pre-equilibrium phase, which lasts less than 1 fm/$c$, are usually entirely ignored \cite{JETSCAPE:2021ehl}. However, recent kinetic theory calculations suggest that these effects are sizable \cite{Das:2015aga,Das:2017dsh,Li:2020kax,Song:2019cqz,Song:2020tfm,Kumar:2021goi}. We are interested in the even earlier pre-equilibrium phase when the medium is not described in terms of quasi-particles, as in a kinetic theory, but rather as a system of chromodynamic fields. 

Within the framework of the Color Glass Condensate (CGC) approach, see {\it e.g.} the review \cite{Gelis:2012ri}, color charges of partons confined in the colliding nuclei act as sources of long wavelength chromodynamic fields which can be treated classically,  because of their large occupation numbers. This non-equilibrium system that exists at very early times is called a glasma \cite{Gelis:2012ri}. It has been argued \cite{Mrowczynski:2017kso,Carrington:2020sww} that this transient phase significantly contributes to the energy loss of hard probes because of its very high energy density. Recently, there has been a lot of interest in the study of the transport properties of this system \cite{Ruggieri:2018rzi,Sun:2019fud,Liu:2019lac,Liu:2020cpj,Khowal:2021zoo,Ipp:2020mjc,Ipp:2020nfu,Boguslavski:2020tqz}.

Our objective is to systematically study the role of the glasma in jet quenching and compare the effect of the pre-equilibrium phase with that of equilibrium quark-gluon plasma evolving hydrodynamically.  We use a Fokker-Planck equation derived in \cite{Mrowczynski:2017kso} whose collision terms encode information about glasma dynamics through correlators of chromodynamic fields. We calculate the relevant correlators using a CGC approach and expand the glasma fields in the proper time $\tau$. Because of the glasma's short lifetime, the proper time can be treated as an expansion parameter. This approach to analytically solve the classical Yang-Mills equations was proposed in \cite{Fries:2006pv}, and developed in \cite{Chen:2015wia}. Using a proper-time expansion, the solutions for the glasma potentials can be found recursively to any order in $\tau$, in terms of the initial gauge potentials.  

In our previous work \cite{Carrington:2020sww} we developed a method to calculate the correlators that enter the Fokker-Planck collision terms using a CGC approach, working at lowest order in the proper-time expansion. In this Letter we present the results from a fifth order calculation, find the radius of convergence of the expansion, and obtain an estimate of the importance of the effect of the glasma on the process of jet quenching. A full account of our study is given in our extensive publication \cite{Carrington:2022bnv}, where we present calculations of the transverse momentum broadening and collisional energy loss in glasma. Their dependencies on time, probe velocity and the values of the parameters that enter CGC calculations are discussed in detail. The regularization schemes that are used and the limitations of our approach are also analyzed. 

We note that our study of hard probes is a part of our bigger project which aims at exploring properties of the glasma. Using a proper-time expansion we have calculated the energy-momentum tensor and obtained from it several observables that describe the evolution of the glasma state  \cite{Carrington:2020ssh,Carrington:2021qvi}.

We use natural units with $c = \hbar = k_B =1$. The indices $\alpha, \beta = 1, 2, 3$ label the Cartesian spatial coordinates while the indices $i, j$ are reserved for transverse coordinates $x, y$. The generators $t^a$ of the SU($N_c$) group are defined through $[t^a,t^b] =  i f^{abc}t^c$ where $f^{abc}$ are the structure constants. The indices $a, b, c = 1, 2, \dots N_c^2 -1$ numerate color components in the adjoint representation. The generators in the fundamental representation are normalized by ${\rm{Tr}}(t^a t^b)=\frac{1}{2} \delta^{ab}$. The generators in the adjoint representation are $T^c_{ab} = i f^{abc}$. Since we study chromodynamics only, the prefix `chromo' is neglected henceforth when referring to chromoelectric or chromomagnetic fields.

The Fokker-Planck equation of hard probes in a plasma populated with strong chromodynamic fields is \cite{Mrowczynski:2017kso}
\be
\label{F-K-eq}
\Big({\cal D} - \nabla_p^\alpha  X^{\alpha\beta}({\bf v}) \nabla_p^\beta - \nabla_p^\alpha  Y^\alpha ({\bf v}) \Big) n(t, {\bf x},{\bf p}) = 0,
\ee
where ${\bf v}={\bf p}/E_{\bf p}$ is the velocity of the hard parton, ${\bf p}$ is its momentum and $E_{\bf p}$ is its energy. The parton's four-momentum is assumed to be on mass-shell. ${\cal D} \equiv  \frac{\partial}{\partial t} + {\bf v}\cdot \nabla $ is the substantial derivative. The tensor  $X^{\alpha\beta}({\bf v})$ is 
\be
\label{X-def}
X^{\alpha\beta}({\bf v}) \equiv
\frac{1}{2N_c} \int_0^t dt' \: {\rm Tr}\big[ \big\langle \mathcal{F}^\alpha(t, {\bf x}) 
\mathcal{F}^\beta \big(t-t', {\bf x}-{\bf v} t'\big) \big\rangle \big] ,
\ee
where $\mathbfcal{F} (t,{\bf x})$ is the Lorentz color force $\mathbfcal{F} (t,{\bf x}) \equiv g \big({\bf E}(t,{\bf x}) + {\bf v} \times {\bf B}(t,{\bf x})\big)$, and $g$ is the coupling constant. The electric ${\bf E}(t,{\bf x})$ and magnetic ${\bf B}(t,{\bf x})$ fields are given in the fundamental representation of the SU($N_c$) group. 

For a non-equilibrium system it is unclear how to derive the vector $Y^\alpha({\bf v})$ in   Eq.~(\ref{F-K-eq}). Therefore, we adopt the plausible assumption that the equilibrium distribution function $n^{\rm eq}({\bf p}) \sim \exp (-E_{\bf p}/T)$ is a solution of the Fokker-Planck equation (\ref{F-K-eq}), where $T$ is the temperature of an equilibrated quark-gluon plasma that has the same energy density as the glasma. This means that  the vector $Y^\alpha({\bf v})$ has to satisfy the relation
\be
\label{XY}
Y^\alpha({\bf v}) = \frac{v^\beta}{T} X^{\alpha\beta} ({\bf v}) .
\ee
We use Eq.~(\ref{XY}) to get $Y^\alpha({\bf v})$ from $X^{\alpha\beta} ({\bf v})$. However, it should be stressed that the vector  $Y^\alpha({\bf v})$ is not needed to obtain $\hat{q}$ which is the subject of this Letter. This vector is required to have the complete Fokker-Planck equation and to compute the collisional energy $dE/dx$ studied in \cite{Carrington:2022bnv}.

The Fokker-Planck equation (\ref{F-K-eq}) is derived in \cite{Mrowczynski:2017kso} under the assumption that the microscopic chromodynamic field and distribution function of hard probes can be decomposed into regular slowly evolving components and rapidly fluctuating ones. The macroscopic field and distribution function are identified with the regular components. The interaction of hard probes with short scale field fluctuations results in Brownian motion. The validity of the Fokker-Planck equation requires that the time scale of field fluctuations, which in the glasma is of the order of the inverse saturation scale $Q_s^{-1}$, is much shorter than the characteristic relaxation time of a hard probe. Since multiple soft interactions are needed to significantly change the momentum of a hard probe  in the glasma, the condition is satisfied.

The field correlators in Eq.~(\ref{X-def}) are non-local and consequently, they are not gauge invariant. The problem can be remedied by inserting a link operator between the two fields, as discussed in Ref.~\cite{Mrowczynski:2017kso}. When the force $\mathbfcal{F}$ is in the adjoint representation, one inserts between the forces in Eq.~(\ref {X-def}) the link $\Omega(t,{\bf x}|t-t',{\bf x}-{\bf v}t')$ defined as
\be
\label{link-def}
\Omega (t_1,{\bf x}_1|t_2,{\bf x}_2) \equiv {\cal P} \exp\Big[ ig \int_{(t_2,{\bf x}_2)}^{(t_1,{\bf x}_1)} ds_\mu A^\mu_c(s) T^c\Big] ,
\ee
where ${\cal P}$ denotes ``left later'' path ordering. Since the analytic calculations become prohibitively difficult with the link operator included, we use the expression in equation (\ref{X-def}). At the end of the Letter we give a quantitative estimate of the effect of neglecting the link operator.

The tensor $X^{\alpha\beta}({\bf v})$ is given by correlators of the fields ${\bf E}$ and ${\bf B}$ in the glasma. When the system under consideration is translationally invariant $X^{\alpha\beta} ({\bf v})$ is independent of the variable ${\bf x}$ which is present on the right side of Eq.~(\ref{X-def}). Since we deal with a system which, in general, is not translationally invariant, we expect a weak dependence of $X^{\alpha\beta} ({\bf v})$ on position which is not explicitly shown on the left side of Eq.~(\ref{X-def}). The tensor $X^{\alpha\beta}({\bf v})$ is expected to saturate at large enough time $t$. This independence occurs due to finite correlation lengths. If the correlator $\big\langle \mathcal{F}^\alpha(t, {\bf x}) \mathcal{F}^\beta (t', {\bf x}') \big\rangle$ vanishes for $|{\bf x}'- {\bf x}| > \lambda_x$ or $|t'- t| > \lambda_t$, the integral (\ref{X-def}) saturates for $ t > \lambda_t$ or  $ t > \lambda_x/v$ with $v \equiv |{\bf v}|$.

We are interested in the momentum broadening coefficient $\hat{q}$ which determines the radiative energy loss of a high-energy parton traversing the medium \cite{Baier:1996sk}. It can be expressed in terms of the tensor $X^{\alpha\beta}({\bf v})$ as \cite{Mrowczynski:2017kso}
\be
\hat{q} =  \frac{2}{v} \Big(\delta^{\alpha\beta} - \frac{v^\alpha v^\beta}{v^2}\Big) X^{\alpha\beta}({\bf v}).
\ee

We consider a collision of two heavy ions moving with the speed of light towards each other along the $z$-axis and colliding at $t=z=0$. The nuclei are assumed to be of infinite extent and homogeneous in the transverse plane. The surface color charge densities of the nuclei are both equal to the constant $\mu$ which we relate to the saturation momentum scale $Q_s$ as $\mu = g^{-4} Q_s^2$. The pre- and post-collision gauge potentials are given by the ansatz \cite{Kovner:1995ts,Kovner:1995ja} 
\begin{align}
\nn
&A^+(x) = \Theta(x^+)\Theta(x^-) x^+ \alpha(\tau, {\bf x}_\perp),  
\\[2mm] \label{ansatz}
&A^-(x) = -\Theta(x^+)\Theta(x^-) x^- \alpha(\tau, {\bf x}_\perp), 
\\[2mm] \nn
&A^i(x) = \Theta(x^+)\Theta(x^-) \alpha_\perp^i(\tau, {\bf x}_\perp)
\\ \nn
&~~~~~~~~ + \Theta(-x^+)\Theta(x^-) \beta_1^i(x^-, {\bf x}_\perp)
\\ \nn
& ~~~~~~~~ + \Theta(x^+)\Theta(-x^-) \beta_2^i(x^+, {\bf x}_\perp) ,
\end{align}
where $x^\pm \equiv (t \pm z)/\sqrt{2}$, $\tau \equiv \sqrt{2 x^+ x^-}$ and $\eta \equiv \frac{1}{2}\ln (x^+/x^-)$. The step functions separate the pre-collision and post-collision regions of spacetime. The potential $\beta^i_1(x^-,{\bf x}_\perp)$ is the pre-collision potential of the right moving nucleus and $\beta^i_2(x^+,{\bf x}_\perp)$ is the pre-collision potential of  the left moving nucleus. The glasma potentials produced after the collision are represented by $\alpha(\tau,{\bf x}_\perp)$ and $\alpha^i_\perp(\tau,{\bf x}_\perp)$ and they are smooth functions in the forward light-cone region. Because of boost invariance, the glasma potentials do not depend on $\eta$.  

To calculate the field correlators, the glasma potentials are expanded in powers of the proper time $\tau$. The Yang-Mills equations are solved recursively so that higher order expansion coefficients are given in terms of the zeroth order glasma potentials, which are written in terms of the pre-collision potentials $\beta^i_1 (x^-,{\bf x}_\perp)$, $\beta^i_2 (x^+,{\bf x}_\perp)$ and their derivatives, using boundary conditions obtained by integrating the Yang-Mills equations across the boundary between the pre-collision and post-collision regions. These relations were originally obtained for infinitely Lorentz contracted nuclei \cite{Kovner:1995ts,Kovner:1995ja}. The boundary conditions we use are \cite{Carrington:2020ssh}
\be
\label{cond1}
\alpha^i_\perp (0,{\bf x}_\perp)  = \lim_{{\rm w}\to 0} 
\big(\beta^i_1 (x^-,{\bf x}_\perp)+\beta^i_2 (x^+,{\bf x}_\perp) \big),
\ee
\be
\label{cond2}
\alpha(0,{\bf x}_\perp)  = -\frac{ig}{2} \lim_{{\rm w}\to 0} 
\big[\beta^i_1 (x^-,{\bf x}_\perp),\beta^i_2 (x^+,{\bf x}_\perp)\big],
\ee
where ${\rm w}$ denotes the longitudinal extent of each nucleus across the light cone, which is introduced for technical reasons 
and taken to zero at the end of the calculation. 

The electric and magnetic fields produced between the receding nuclei are also expanded in $\tau$
\be
\label{E-tau}
{\bf E}(\tau,{\bf x}_\perp,\eta) = {\bf E}^{(0)}({\bf x}_\perp) 
+ \tau {\bf E}^{(1)}({\bf x}_\perp,\eta) + \dots,
\ee
\be
\label{B-tau}
{\bf B} (\tau,{\bf x}_\perp,\eta)= {\bf B}^{(0)}({\bf x}_\perp) 
+ \tau {\bf B}^{(1)}({\bf x}_\perp,\eta) + \dots. 
\ee
The initial or zeroth order fields are parallel to the beam direction and are given by
\ba
\label{E0}
&&E^{(0)}({\bf x}_\perp) = - 2 \alpha^{(0)}({\bf x}_\perp) , 
\\[2mm]
\nn
&&B^{(0)} ({\bf x}_\perp) =
\partial^y \alpha^{x(0)}_\perp({\bf x}_\perp) 
- \partial^x \alpha^{y(0)}_\perp({\bf x}_\perp)
\\[1mm] \label{B0}
&&~~~~~~~~~~~~ -ig [\alpha^{y(0)}_\perp({\bf x}_\perp),\alpha^{x(0)}_\perp({\bf x}_\perp)].
\ea

We need to calculate correlators of pre-collision potentials of the form
\be
\label{corr-gen}
\langle \beta^i_{a1} \beta^j_{b1} \dots \beta^k_{c2} \beta^l_{d2} \dots  \rangle.
\ee
We use the Glasma Graph approximation \cite{Lappi:2015vta} which means that Wick's theorem, which properly speaking should only be applied to a product of color sources, is used on a product of pre-collision potentials. Correlators of an odd number of the potentials vanish and correlators of an even number of potentials can be expressed as products of pairs of pre-collision correlators. One also assumes that potentials from different nuclei are uncorrelated, so that we only need to calculate the correlator
\be
\label{field-corr}
\delta_{ab} B^{ij}({\bf x}_\perp,{\bf y}_\perp) = \lim_{{\rm w}\to 0} \langle \beta^i_{a}(x^\mp,{\bf x}_\perp) \beta^j_{b}(y^\mp,{\bf y}_\perp)\rangle,
\ee
where the upper/lower sign on the light-cone variables corresponds to the first/second ion. A detailed derivation of the correlator (\ref{field-corr}), following the presentation in \cite{Chen:2015wia}, can be found in \cite{Carrington:2020ssh}.

\begin{figure*}[t]  
\begin{minipage}{79mm} 
\centering
\includegraphics[scale=0.33]{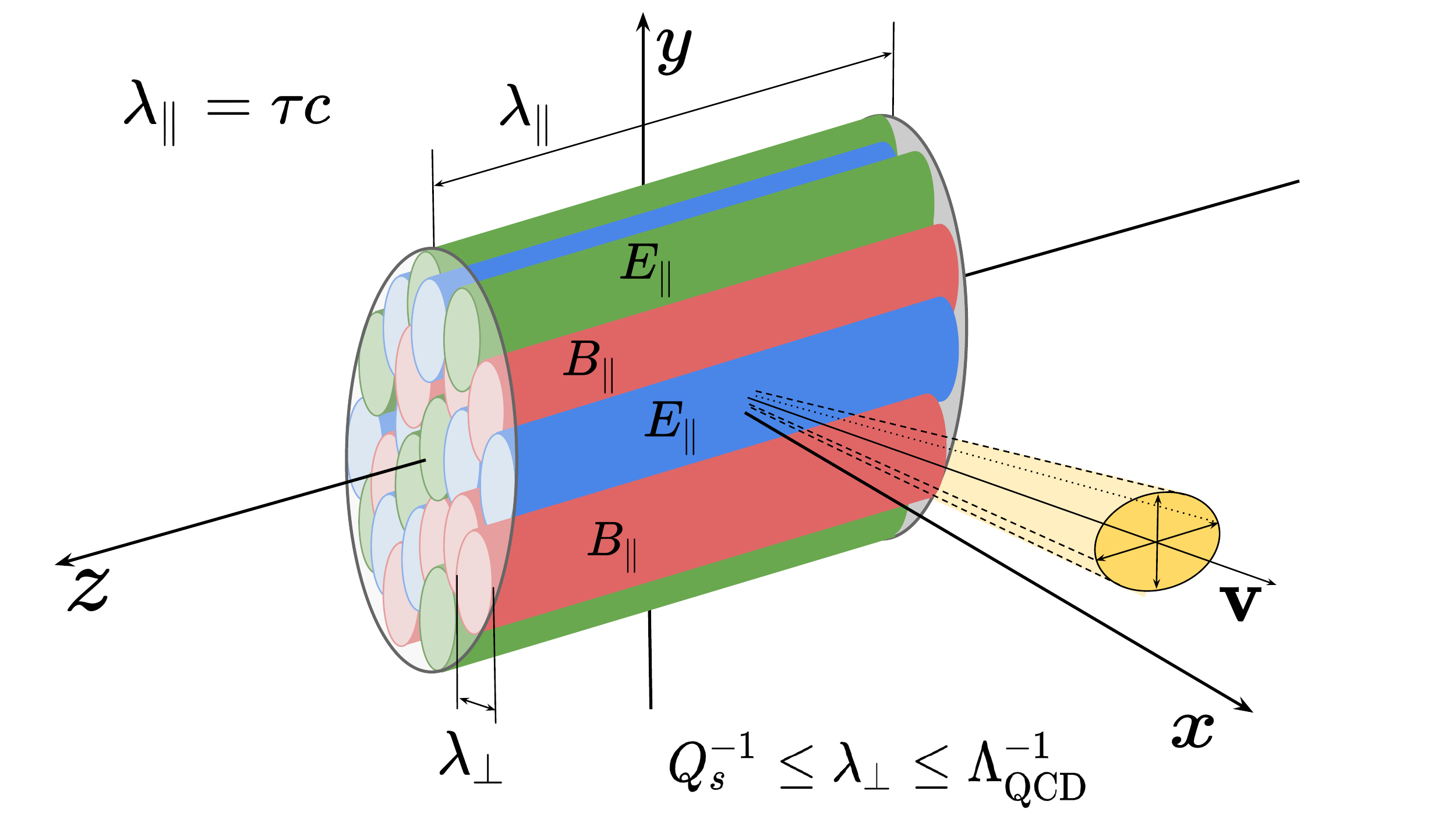}  
\vspace{-3mm}
\caption{Cartoon of the zeroth order glasma fields and a probe moving mostly transverse to the collision axis.} 
\label{tubes}
\end{minipage}
\hspace{5mm}
\begin{minipage}{79mm}
\centering
\includegraphics[scale=0.19]{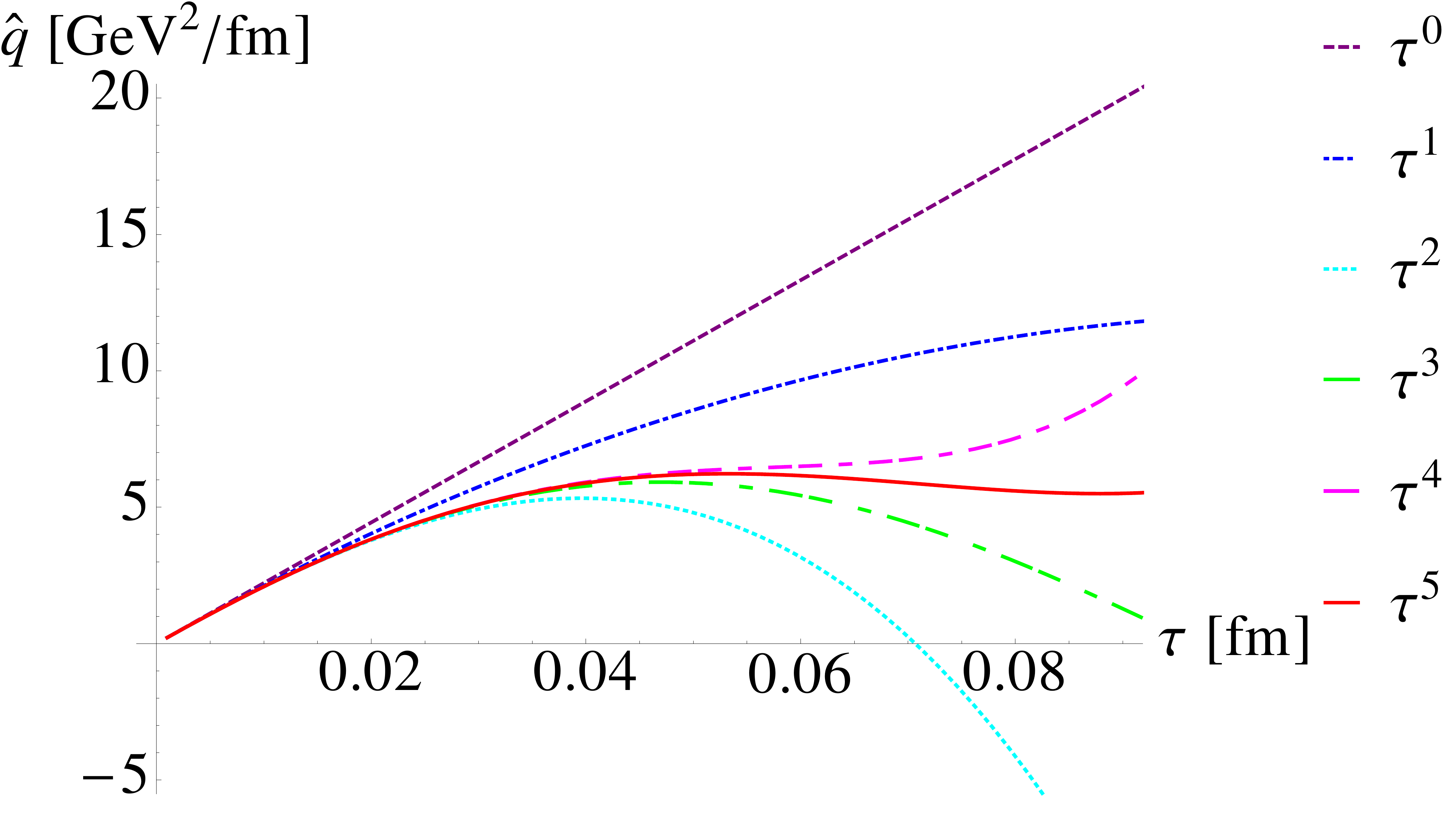} 
\vspace{-1mm}
\caption{Time evolution of $\hat{q}$ at different orders of the proper-time expansion.}
\label{t-coeff-cum} 
\end{minipage}
\end{figure*} 

The correlator (\ref{field-corr}) takes the form
\be
\label{Ai0-Aj0-C1-C2}
B^{ij}(r)
= \delta^{ij}  C_1 (r) - \hat{r}^i \hat{r}^j C_2 (r) ,
\ee
where $r \equiv|{\bf x}_\perp - {\bf y}_\perp|$ and $\hat{r}^i \equiv r^i/r$. The functions $C_1(r)$ and $C_2(r)$ are 
\ba
\label{C1-def}
C_1(r) \equiv  \frac{m^2 K_0 (mr)}{g^2 N_c  \big( mr K_1(mr) - 1\big) } \, M(r) ,
\\[2mm]
\label{C2-def}
C_2(r) \equiv  \frac{m^3 r \, K_1(mr) }{g^2 N_c  \big( mr K_1(mr) - 1\big) } \, M(r) ,
\ea
where 
\be
M(r) \equiv \exp\bigg[\frac{g^4 N_c \mu \big( mr K_1(mr) - 1\big)  }{4\pi m^2 }\bigg] -1 ,
\ee
$K_0(x)$ and $K_1(x)$ are the Macdonald functions and $m$ is an infrared regulator which is identified with $\Lambda_{\textrm{QCD}}$, so that $m \approx \Lambda_{\textrm{QCD}} \approx 200$ MeV. This choice naturally encodes the behavior of confinement, since it ensures that the color charges in a nucleon are neutralized at the length scale which coincides with $\Lambda_{\textrm{QCD}}^{-1}$. The function $C_1(r)$ diverges logarithmically as $r \to 0$ and hence the correlator (\ref{Ai0-Aj0-C1-C2}) has to be regularized. Any function of $r$ for $r < Q_s^{-1}$  is replaced by its constant value taken at $Q_s^{-1}$. In Ref.~\cite{Carrington:2022bnv} we show that our results depend only weakly on this regularization.

The correlators of electric and magnetic fields can be expressed in terms of the functions (\ref{C1-def}) and (\ref{C2-def}). At zeroth order the correlators are 
\begin{align}
\nn
& \langle E^{(0)}_a({\bf x}_\perp) \, E^{(0)}_b({\bf y}_\perp)\rangle  = g^2 N_c \delta^{ab} 
\\ \label{EzEz-corr}
& ~~~~~~~~ \times
\big( 2 C_1^2 (r) - 2 C_1 (r) \, C_2 (r) + C_2^2 (r) \big) ,
\\[2mm] \nn
& \langle B^{(0)}_a({\bf x}_\perp) \, B^{(0)}_b({\bf y}_\perp)\rangle  = g^2 N_c \delta^{ab} 
\\ \label{BzBz-corr}
& ~~~~~~~~ \times
\big( 2 C_1^2 (r) - 2 C_1 (r) \, C_2 (r) \big),
\\[2mm]
\label{EB}
& \langle E^{(0)}_a({\bf x}_\perp) \, B^{(0)}_b({\bf y}_\perp)\rangle  = 0 .
\end{align}
Higher order correlators are given by similar expressions involving the functions $C_1(r)$ and $C_2(r)$ and their derivatives. 
The integrand on the right side of Eq.~(\ref{X-def}) also involves correlators of pairs of fields at the same point, and we treat these as two-point correlators in the limit $r\to 0$. Combining the results for all correlators, we obtain an analytic expression for the tensor $X^{\alpha\beta}({\bf v})$ up to order $\tau^5$.

The proper-time expansion at the lowest orders in $\tau$ is qualitatively similar to the semi-analytic weak-field approximation which was discussed in \cite{Ipp:2020nfu} and used to check the correctness of the fully non-perturbative lattice code. The weak-field approximation is based on the assumption that the charge densities of colliding nuclei are small. Consequently, the chromodynamic fields generated in the collision are weak, effectively Abelian, and the glasma temporal evolution can be treated as a perturbation of the initial configuration with electric and magnetic fields along the beam direction. Using a proper-time expansion, which does not assume that the fields are weak, our study shows that the initial field configuration lasts for a very short time and evolves rapidly.

An idealized picture of a probe going through the glasma at very early times is shown in Fig.~\ref{tubes}, where the glasma fields at zeroth order in the $\tau$ expansion are represented by colored flux tubes of electric and magnetic fields which are longitudinal and static. Because experiments at RHIC and the LHC focus on hard probes from the mid-rapidity region, $-1 < y < 1$, we are interested in probes moving mostly perpendicularly to the beam axis. There are two qualitatively different correlation lengths, which we denote $\lambda_\parallel$ and $\lambda_\perp$. The longitudinal correlation length $\lambda_\parallel$ is roughly the distance between the nuclei and can be identified with the proper time $\tau$. The transverse correlation length $\lambda_\perp$ can be inferred from the correlators (\ref{EzEz-corr}) and (\ref{BzBz-corr}). Qualitatively it obeys $Q_s^{-1} \leq\lambda_\perp \leq \Lambda^{-1}_{\rm QCD}$. Beyond zeroth order, transverse electric and magnetic fields develop and rapidly change in time. The simple picture in Fig.~\ref{tubes} does not accurately describe the glasma except at very early times.

The momentum broadening $\hat{q}$ is built up during the time that the probe spends within the domain of correlated fields. At zeroth order, this time is determined by the transverse correlation length and the orientation and magnitude of the probe's velocity. The coefficient $\hat{q}$ saturates when the probe leaves the region of correlated fields. We show below that saturation is reached before the $\tau$ expansion breaks down, which proves that the proper-time expansion, at the order we work at, is consistent with the approximations that are used in the Fokker-Planck approach. 

In addition to determining how long a probe spends in the region of correlated fields, the probe's velocity influences momentum broadening through the Lorentz force which enters Eq.~(\ref{X-def}). Since the electric and magnetic fields are mostly along the beam direction (see Eq.~(\ref{B0})), the momentum broadening is maximal when the probe moves transversely to the collision axis. In this Letter we focus on this case. The dependence of $\hat{q}$ on the orientation of the probe velocity is discussed in \cite{Carrington:2022bnv}. 

The momentum broadening for a quark moving with the speed of light perpendicularly to the beam axis is presented in Fig.~\ref{t-coeff-cum} as a function of $\tau$. The dependence of $\hat q$ on the order in the $\tau$ expansion is shown. We observe that taking into account higher and higher order contributions clearly extends the range of validity of the $\tau$ expansion, which can be estimated from the largest value of $\tau$ for which the result at a given order agrees reasonably well with the result at the previous order. At very early times, all orders of the $\tau$ expansion agree well. When all terms up to order $\tau^5$ are included, the time evolution of $\hat q$ shows initial growth, and then flattening, followed by more rapid growth. The region $0.03 ~ \text{fm} \lesssim \tau \lesssim 0.07$ fm where $\hat q$ flattens indicates the saturation of the result. The rapid increase of $\hat q$ at later times reflects the break down of the proper-time expansion. At order $\tau^5$, the highest value of $\hat q$ that is obtained before the proper-time expansion breaks down is around 6 GeV$^2$/fm. The results in Fig.~\ref{t-coeff-cum} are obtained using $N_c=3$, $g=1$, $Q_s=2$ GeV and $m=0.2$ GeV. The dependence of $\hat{q}$ on the values of $Q_s$ and $m$ is discussed in \cite{Carrington:2022bnv}. We note that the momentum broadening of a hard gluon can be obtained by multiplying the quark value of $\hat{q}$ by the color factor $9/4$. 

We have found that $\hat q$ in the glasma reaches the maximal value of $\hat{q} \approx 6 ~ {\rm GeV^2/fm}$ at $\tau = 0.06$ fm. The coefficient $\hat{q}$ has also been calculated using real time CGC simulations \cite{Ipp:2020nfu}, and shows similar growth and a comparable magnitude over the same range of proper times. Even though the result we have obtained for the momentum broadening coefficient is very large, one wonders whether the glasma significantly contributes to the total momentum broadening that the probe experiences when it moves through the system. The point is that the pre-equilibrium phase exists for less than 1 fm.

The value of $\hat{q}$ in equilibrium quark-gluon plasma, which is inferred from experimental data by the JETSCAPE Collaboration \cite{JETSCAPE:2021ehl}, is $2 < \hat{q}/T^3 < 4$ where $T$ is the plasma temperature. The expression is valid for a hard quark of $p_T > 40$ GeV. We take $\hat{q} = 3T^3$ for our further discussion. Since the temperature of the plasma produced at the LHC evolves from roughly 450 to 150 MeV \cite{Shen:2011eg}, the momentum broadening varies from $\hat{q} \approx 1.0~{\rm GeV^2/fm}$ to $\hat{q} \approx 0.05~{\rm GeV^2/fm}$, which are significantly smaller than the glasma value $\hat{q} \approx 6 ~ {\rm GeV^2/fm}$. 

\begin{figure}[b]
\centering
\includegraphics[scale=0.23]{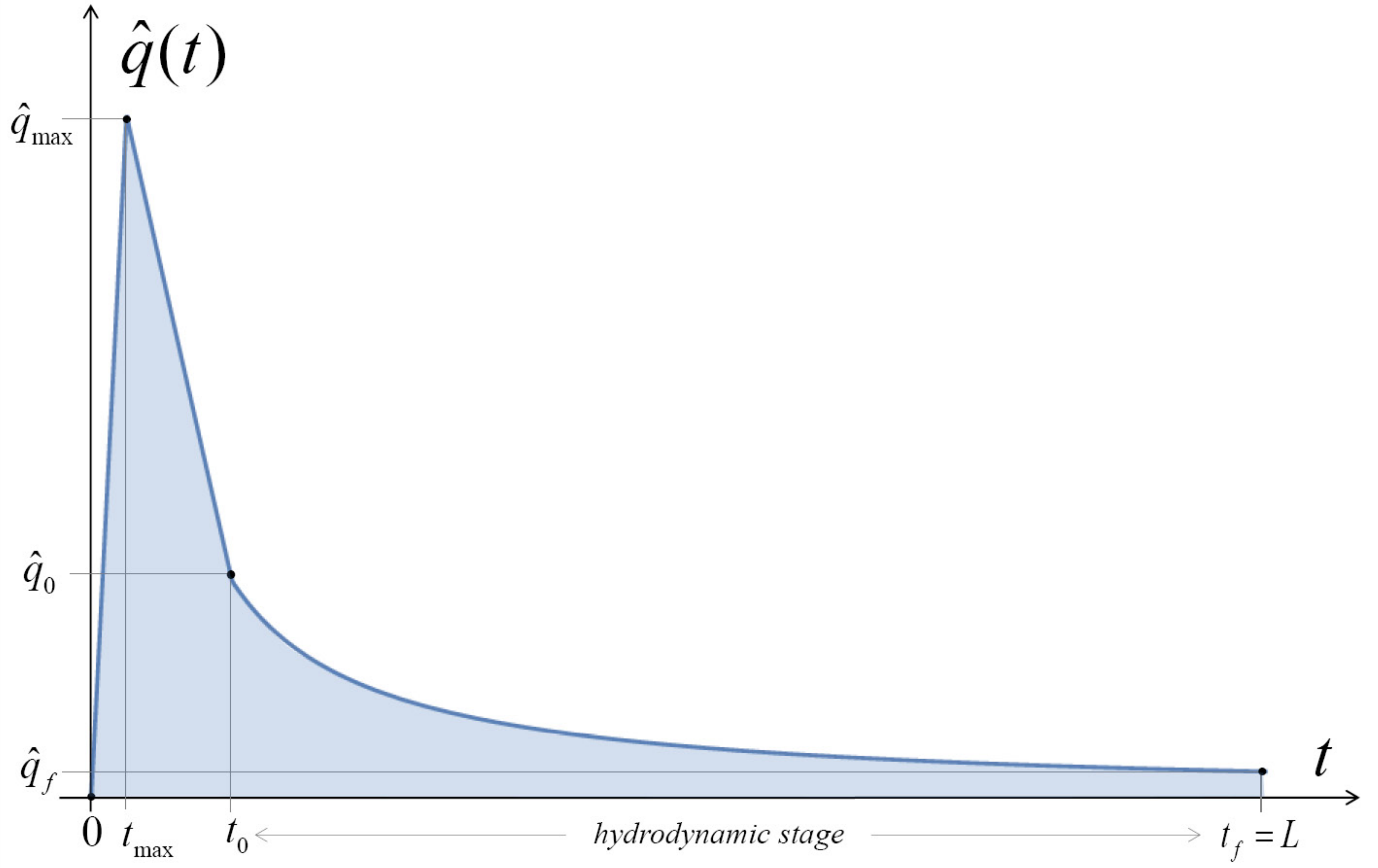}
\caption{Schematic temporal evolution of $\hat{q}(t)$.} 
\label{Fig-qhat-time}
\end{figure}

The radiative energy loss per unit length of a high-energy parton traversing a medium of length $L$ is proportional to the total accumulated transverse momentum broadening  $\Delta p_T^2$ \cite{Baier:1996sk}. In case of a static medium we have $\Delta p_T^2 = \hat{q}L$. When the plasma is not static and $\hat{q}$ is time dependent, the momentum broadening is $\Delta p_T^2 = \int_0^L dt \, \hat{q}(t)$ where the probe is assumed to move with the speed of light.  

The time dependence of the momentum broadening throughout the whole history of the deconfined matter produced in a relativistic heavy-ion collision is shown schematically in Fig.~\ref{Fig-qhat-time}. There is rapid growth of $\hat{q}(t)$ in the glasma phase until a maximal value $\hat{q}_{\rm max} \approx 6~{\rm GeV^2/fm}$  is reached at $t_{\rm max} \approx 0.06$ fm.  At later times $\hat{q}(t)$ decreases to a value of $\hat{q}_0$ at $t_0$ (we explain below how to estimate these values). From this point on, we have equilibrated quark-gluon plasma which expands hydrodynamically \cite{JETSCAPE:2021ehl}. The saturation region seen in Fig.~\ref{Fig-qhat-time} is not visible in Fig.~\ref{t-coeff-cum} because of the difference in the time scales in the two figures. The drop of $\hat q$ between $t_{\rm max}$ and $t_0$ is not captured by our calculation but it is  covered by the simulations of Ref.~\cite{Ipp:2020nfu}.

Assuming that the expansion occurs according to ideal one-dimensional boost-invariant hydrodynamics, the temperature decreases as $T = T_0 (t_0/t)^{1/3}$. Consequently, $\hat{q}$ depends on time as 
\be
\label{qhat-vs-time}
\hat{q}(t) = 3 T_0^3 \, \frac{t_0}{t} = \hat{q}_0 \frac{t_0}{t}.
\ee
Taking the value  $t_0 = 0.6$ fm \cite{JETSCAPE:2021ehl} and $T_0=0.45$ GeV \cite{Shen:2011eg} gives $\hat{q}_0=1.4$ GeV$^2$/fm. 
The equilibrium contribution to $\Delta p_T^2$ is
\be
\Delta p_T^2\big|^{\rm eq}  \equiv \int_{t_0}^L dt \, \hat{q}(t) 
= 3 T_0^3 \, t_0 \, \ln \frac{L}{t_0} .
\ee

Using linear interpolation between the points $\hat{q}(0)=0$, $\hat{q}(t_{\rm max})=\hat{q}_{\rm max}$, and $\hat{q}(t_0)=\hat{q}_0$, one finds the following non-equilibrium contribution to $\Delta p_T^2$
\be
\Delta p_T^2\big|^{\rm neq} \equiv  \int_0^{t_0} dt \, \hat{q}(t) 
= \frac{1}{2}\hat{q}_{\rm max} t_0 +  \frac{1}{2}\hat{q}_0(t_0 - t_{\rm max}) .
\ee
The values $\hat{q}_{\rm max} = 6~{\rm GeV^2/fm}$, $t_{\rm max} = 0.06$ fm, $t_0 = 0.6$ fm 
and $\hat q_0 = 1.4$ GeV$^2$/fm give
\be
\frac{\Delta p_T^2\big|^{\rm neq}}{\Delta p_T^2\big|^{\rm eq}} = 0.93
\ee
for a path of length $L = 10$ fm. If the value of $\hat{q}_{\rm max}$ is increased, or the initial temperature $T_0$ is reduced, or the path $L$ is shortened, the non-equilibrium phase gives a contribution to the radiative energy loss which is even bigger than the equilibrium one. 

We note that if we simply extrapolate the equilibrium $\hat{q}(t)$ to $t_{\rm max}=0.06$ fm using Eq.~(\ref{qhat-vs-time}) we get $\hat{q}(t_{\rm max}) \approx 14~{\rm GeV^2/fm}$ which is significantly bigger than the glasma value of $\hat{q}(t_{\rm max}) \approx 6 ~ {\rm GeV^2/fm}$ we found. This shows that an equilibrium quark-gluon plasma with an energy density as high as that of the glasma would be similarly opaque to hard probes. 

Finally, we return to the issue of the gauge dependence of our results due to the absence of the link operator (\ref{link-def}) in the field correlators. We compute the statistical ensemble average of the operator (\ref{link-def}). The crucial observation is that the magnitude of the tensor (\ref{X-def}) is saturated after a time of approximately 0.06 fm. Consequently, the time interval covered by the link operator is very short and thus the operator can be approximated by the first three terms of the expansion of the exponential function in (\ref{link-def}). Since the ensemble average of a single potential vanishes, we obtain
\begin{align}
\label{ave-Omega-approx}
&\frac{1}{N_c^2-1} \,\big\langle \Omega (t,{\bf x}|t-t', {\bf x}-{\bf v} t')\big\rangle
\\[2mm] \nn
& ~~~~
 = 1 - \frac{g^2 N_c}{2(N_c^2-1)}
\big\langle A_a^\mu(x)  \, A_a^\nu(x) \big\rangle \Delta s_\mu \Delta s_\nu .
\end{align}

We compute the expression (\ref{ave-Omega-approx}) in the zeroth order of the proper time expansion of glasma which allows one to express the gauge potential $A^\mu$ through the pre-collision potentials. Using the correlator (\ref{field-corr}) and assuming that the velocity ${\bf v}$ of hard probe is perpendicular to the collision axis, one obtains
\begin{align}
\label{ave-Omega-approx-final}
& \frac{1}{N_c^2-1} \,\big\langle \Omega (t,{\bf x}|t-t', {\bf x}-{\bf v} t')\big\rangle 
\\ \nn
&~~~~
= 1 -  \frac{N_c Q_s^2}{8\pi} 
\bigg[ \ln\Big(\frac{Q_s^2}{m^2} +1 \Big) -  \frac{Q_s^2}{Q_s^2 + m^2} \bigg]  {\bf v}^2 t'^2 .
\end{align}
With $N_c = 3$, $Q_s=2$~GeV, $m=200$~MeV, ${\bf v}^2 = 1$ and $t' = 0.06$~fm, we find 
\be
\label{ave-link-final}
1 - \frac{1}{N_c^2-1} \,\big\langle \Omega (t,{\bf x}|t-t', {\bf x}-{\bf v} t')\big\rangle = 0.16 .
\ee
This estimate most likely provides an upper limit, for two reasons. First, we have computed the potential correlator at the zeroth order of the proper-time expansion which corresponds to the strongest fields. Second, the correlator (\ref{ave-Omega-approx}) has been taken at the minimal distance where its size is maximal.

Equation (\ref{ave-link-final}) shows that the ensemble average of the link operator per color degree of freedom differs from unity by 0.16, which is not a big number. This result supports the idea that neglecting the link operator in the collision term of the Fokker-Planck equation does not invalidate our results.  

In conclusion, our approach, which is presented in detail in \cite{Carrington:2022bnv}, provides a reliable estimate of momentum broadening in glasma. Our calculation gives a value of $\hat{q}$ that is several ${\rm GeV^2/fm}$, which is much larger than equilibrium values, and produces accumulated transverse momentum broadening of the same order as the contribution from the equilibrium phase. Our results therefore indicate that the transient glasma phase plays an important role in the jet quenching. This conclusion is significant because it contradicts previous beliefs that the contribution to momentum broadening from the glasma phase can be safely neglected. 

\vspace{3mm}

This work was partially supported by the National Science Centre, Poland under grant 2018/29/B/ST2/00646, and by the Natural Sciences and Engineering Research Council of Canada grant 2017-00028.

\end{document}